\documentclass[conference]{IEEEtran}
\author{\IEEEauthorblockN{Benjamin Dowling}
	\IEEEauthorblockA{King's College London\\
		benjamin.dowling@kcl.ac.uk}
	\and
	\IEEEauthorblockN{Britta Hale}
	\IEEEauthorblockA{Naval Postgraduate School\\
		britta.hale@nps.edu}
	\and
	\IEEEauthorblockN{Xisen Tian}
	\IEEEauthorblockA{Naval Postgraduate School\\
		xisen.tian1@nps.edu}
        \and
	\IEEEauthorblockN{Bhagya Wimalasiri}
	\IEEEauthorblockA{University of Sheffield\\
		b.m.wimalasiri@sheffield.ac.uk}
  }

\title{Key Establishment in the Space Environment}

\usepackage{adjustbox}
\usepackage{amsmath}
\usepackage{hyperref} 
\usepackage{cleveref}
\usepackage{url}
\usepackage[mode=buildnew]{standalone}
\usepackage{subcaption} 
\usepackage{tikz}
\usepackage{xspace}
\usetikzlibrary{arrows, shadows, positioning}
\usepackage{stfloats}

\newcommand{\xt}[1]{{\color{orange!80!red} XT: #1}}
\mathchardef\mhyphen="2D

\renewcommand{\O}[1]{\ensuremath{\mathsf{#1}}}

\ifdefined\adversary
\renewcommand{\adversary}{\ensuremath{\mathcal{A}}}
\else
\newcommand{\adversary}{\ensuremath{\mathcal{A}}}
\fi

\newcommand{\PROCflag}{\ensuremath{\mathsf{\PROCflag}}}
\definecolor{mBlue}{HTML}{22a1dc}
\definecolor{mGreen}{HTML}{86cd30}
\definecolor{mLightBrown}{HTML}{e68a00}

\newcommand{\linkgame}[2]{\hyperref[#1]{G#2}}

\newcounter{Bdversary}

\newcommand{\bigO}{\ensuremath{\textsf{O}}\xspace}

\begin{document}

\maketitle

% Provide the keywords *before* the abstract
% When keywords contain macros provide the text version as the optional argument
\newcommand{\Dirac}{Dirac}
% \keywords[Dirac delta function, unit impulse]{\Dirac~$\delta$ function, unit impulse}

% Provide the abstract of your paper

\begin{abstract}
  As reliance on space systems continues to increase, so does the need to ensure security for them. However, public work in space standards have struggled with defining security protocols that are well tailored to the domain and its risks. 
  In this work, we investigate various space networking paradigms and security approaches, and identify trade-offs and gaps. Furthermore, we describe potential existing security protocol approaches that fit well into the space network paradigm in terms of both functionality and security. 
  Finally, we establish future directions for enabling strong security for space communication.
\end{abstract}
  % In space, no-one can hear you communicate. Thus, establishing mechanisms to facilitate secure communications within the increasingly active space domain is a priority. 
    % terrestrial network paradigms, and determine which fits best within the constraints of the space setting. 

% A separate text-only abstract must be supplied in your final version.
% This will be used for web pages and indexing and should not contain macros.
% \begin{textabstract}
   % TODO
% \end{textabstract}
% The content of the paper starts here
\section{Introduction}
%Motivation, why security is important in space and was not a focus before, why it is challenging to get keys, and how there is a gap between space engineering and cryptography, the latter of which often does not account for the nuances of space requirements, which may not be voiced. 

% Not designed for security originally; nation state actors at play; laws restricting space comms / crypto export; long acquisitions and deployment timelines; 

For the goal of scientific exploration, space systems have often been considered low-value from a  cyber-threat standpoint,  and virtually inaccessible. This led to early communication security for space systems being overlooked as largely irrelevant
% as a solved problem. 
% With so much red tape (e.g. export controls) in the past, 
-- after all, what value would there be in hacking a Mars rover?
The security that was considered was relatively primitive, based on symmetric key approaches such as encryption only, without key establishment~\cite{NASAstandard,L3Harris}. 
% The Global Navigation Satellite System (GNSS), which comprises the Global Positioning System (GPS) in the U.S., Galileo in the E.U., Global Navigation Satellite System (GLONASS) in Russia, and BeiDou Navigation Satellite System (BDS) in China, was considered the rare practical use case for space systems for the average user, and even there security threats posed little risk to an average user. 
As such, it is unsurprising that there has been limited development in space communications security over the last several decades in comparison to the immense body of literature designing and analyzing protocols for terrestrial networks. 
However, space systems are increasingly prevalent in industrial use and even relied upon for daily mundane tasks. 
SpaceX's revolutionary reusable rockets came to market in the 2010s~\cite{maidenberg_spacex_2024} and the proliferation of near-space satellites as an Internet technology has revolutionized use of non-terrestrial platforms and possibilities. 
% Since then, proliferated constellations, r
Space based Internet providers~\cite{starlink}, tele-health~\cite{asrar_can_2021}, space tourism~\cite{cookson_will_2024}, astroid mining~\cite{fernholz_astroforge_2024} and a host of other ventures have developed that continue to expand reliance on space -- and its security -- into people's daily lives~\cite{Menez_exploring_2022}.
% and industrial possibilities, and reliance on it continues to grow. 
Now, everything from banking information to critical infrastructure management flows through space connections. 
Public safety, health, financial transactions are all high-value targets and motivate attacks against space communications~\cite{hughes_satellites_2024}. 
Space systems now require that which has never before been an intrinsic goal: secure channel establishment. 
% Security must now catch up with scaling and federation requirements levied by the new space race. 
 %Additionally, the information carried by devices in space (e.g. military sensors, GPS, scientific devices) had low return on investment (ROI) in terms of immediate value to risk/effort involved to hackers other than those employed by governments. Furthermore, publications relating to space communications and cryptography were banned under International Traffic and Arms Regulations (ITAR) stunting research on the matter \xt{cite}. 

Secure channels are frequently defined by encryption and authentication, ensuring that the data sent and received remains private and unaltered. Such cryptographic functions require secret keys (either symmetric or asymmetric).
% 
% , which raises a fundamental question:
% \begin{center}
%     \emph{What key establishment method is appropriate \\ for current space architecture approaches?}
% \end{center}
% 
% As such,  common approaches used in terrestrial networks can be enticing.
% Firstly, 
Some initial security approaches applied manually-installed, pre-shared keys which are among approaches recommended by the Consultative Committee for Space Data Systems (CCSDS) \cite{rfc9173, SDLS_EP_2020}. This approach has naturally limited scaling, not to mention exceedingly long vulnerability windows due to the typical service lifetime of a system and infeasibility of updating systems after launch. 
% 
% Since space system communication, whether between satellites or in connection with ground terminals, mirrors the networking paradigm of routers, except in this case the routers are non-terrestrial. 
% 
Consequently, other approaches have attempted to project terrestrial network protocols onto space systems~\cite{ccsds_security_2012, huitema-quic-in-space-00}, which would allow for ad-hoc session establishment. Yet, space is not a terrestrial environment: delay times in transmission imply an even higher delay cost from key establishment protocols that require 1.5-2 round trips. If such protocols operate point-to-point (e.g., link layer), they introduce a further risk of data being in the clear at every network `hop', i.e., space or ground terminal. 
% 
% Traditional security for space systems focused on using manually installed pre-shared keys for link-layer bulk encryption\cite{}. 
%or making link layer encryption efficient through hardware bulk encryption devices and transmission security (e.g. RF modulation for low probability of intercept) all based on pre-shared keys (PSKs) manually installed prior to launch. 
% Even though public key cryptographic systems have existed since the 1970s, pre-shared symmetric keys were favored due to the small number of space systems in service, their decades long development and service lifespans, and the infeasibility of updating systems after launch. Particularly for deep space, just-in-time and interactive public key protocols used to perform key exchange such as TLS are ill-suited due to minutes-long delay times (e.g. Earth to Mars). 
Furthermore, packet dropping and link unavailability could incur retransmission attempts. For protocols that require an entire new session establishment after failure,  further precious power and bandwidth resources are wasted. In addition, the threat of quantum computing towards current cryptographic key establishment approaches necessitates use of new cryptographic primitives for key establishment and authentication, which frequently come with larger ciphertexts and public keys.
This raises a fundamental question:
\begin{center}
    \emph{What key establishment method is appropriate \\ for current space architecture approaches?}
\end{center}

Adaptation of key exchange protocols to their use case is not a new endeavor. Key exchange for the Internet of Things (IoT) has been customized for similar reasons~\cite{troncoso_bluetooth_2021, 10.1145/3384217.3385617, noauthor_zigbee_2023, noauthor_bluetooth_2024}, due to use case and environment requirements.
Like the IoT setting, simply ``gluing'' key establishment protocols designed for Internet use cases into the space setting results in suboptimal operation at best. 
Internet protocols can rely on consistent, low-latency bidirectional connections which simply does not apply to the space setting. 
Cellular communication, such as those standardized by 3GPP~\cite{sultan_5g_2022} for 5G, similarly does not rely on common Internet communication key establishment protocols. Thus, key establishment protocols for space systems should consider the unique characteristics of the setting, including delays, relay reliance, 
% packet drop, and relay environment.
intermittent link availability and high latency, properties that are further exacerbated by distance, space weather, and system power limitations.

In this work, we look at two current approaches to space networking and the role of key establishment in each. We summarize  
Delay Tolerant Networking (DTN),  
% using the Bundle Protocol and open questions on key establishment therein, 
designed for deep space interplanetary communications centered on the store-and-forward paradigm, 
using the Bundle Protocol, and open questions on key establishment therein. 
% attacks the communications problem, there is still a gap in security and efficiency due to its the reliance on manually pre-installed keys. 
We also discuss current trends using IP-based protocols such as QUIC within the space setting, which is widely adopted for internet applications like Facebook~\cite{joras_how_2020}, YouTube \cite{noauthor_github_nodate}, and Instagram\cite{joras_how_2020}, and observe ill-fitting characteristics of its key establishment approach for space systems.  Finally, we discuss Continuous Key Agreement (CKA) protocols -- a class of key establishment protocols suited to space environments. These have been previously proposed, being apparently first publicly proposed in 2021~\cite{talk} and later also described in~\cite{talk2,talk3,Bader}. However, although prior work describes the benefits of CKAs in the space setting, how it can be incorporated into current efforts for space protocol stacks has not been described. This work takes a closer look at that gap, showing how CKAs can be incorporated into \emph{either} the DTN or IP-based transport stacks for ensuring secure channels as a unified and functionally appropriate security approach.
There are various uses or protocols related to satellites and other space systems that drive work on DTN and QUIC. We observe for the first time how integration of RFC9420~\cite{rfc9420}, a standardized CKA by the Internet Engineering Task Force (IETF), can solve a key establishment challenges and open issues in approaches in both of these cases. We also discuss how protocol changes to QUIC can be performed to incorporate such a CKA.

\section{Space Networking Stacks}

%Bundle Protocol Setting and Paradigms
The 
% Internet Engineering Task Force (IETF) 
IETF and the Consultative Committee on Space Data Systems (CCSDS) have investigated and made efforts towards the use of DTN protocols in space. There is also growing enthusiasm to use the IP stack architectures as an alternative to or in conjunction with the DTN stack \cite{9172509, huitema-quic-in-space-00}. Supporting arguments of either approach depend on the particular space mission and their locations in space: \textit{near-earth}, \textit{lunar}, or \textit{deep-space} \cite{manning_frequently_2023, consultative_committee_for_space_data_systems_rationale_2010}.  We briefly summarize this context  to lay the ground work for discussion of security approaches. However, recommendations for selection of or use-case preferences among the different fundamental networking stack approaches is out of scope of this work.

\textit{Near-earth.} Communications in Low Earth Orbit (LEO) and Geostationary Orbit (GEO) have the advantage of relatively low round-trip times, on the order of 20 milliseconds (ms) to 250 ms respectively. The relative lower delay makes it potentially feasible to use  
% makes it appealing to use 
traditional IP protocols which have the benefit of widespread implementation and performance characteristics in low latency settings. However, even at these distances link availability is still susceptible to the effects of atmospheric conditions (e.g. rain, clouds, scintillation) and orbit dwell times (e.g. context switching in  LEOs and downtime from perigee of Highly Elliptical Orbits). To ameliorate these constraints, discussions on IP stack use in near-earth missions generally include careful pre-planning (i.e. network topology designs) and additional infrastructure to close the gaps in connectivity (e.g. more ground station transceivers or proliferated mega-constellation networks) \cite{10.1145/3359989.3365407, DELPORTILLO2019123}. Prior research suggests that DTN protocols begin to eclipse IP in performance (i.e., better goodput \cite{floyd_congestion_2000}) when subjected to round-trip times longer than 200ms \cite{6533337}.

\textit{Deep Space.} In deep space (i.e. beyond lunar), the near-earth concerns are amplified to the point where TCP/IP protocols alone have received skepticism on insufficiency \cite{consultative_committee_for_space_data_systems_rationale_2010}. The transmission round-trip time from Earth to Mars, for example, is on the order of just under one minute to 23 minutes depending on the orbital positions of the two systems \cite{noauthor_mars_2023}. Further, systems that operate in deep space must be even more judicious with power and resource conservation: their replenishment is cost prohibitive after launch. %The cryptographic overhead caused by stateless security protocols such as TLS using handshake negotiations could result in self-imposed denial of service. \xt{CONTINUE} %The DTN architecture was created to account for the unique environmental challenges to communications that plague deep space settings which are incompatible with interactive or reactionary protocols found in terrestrial network architectures. Whereas traditional TCP/IP protocols assume continuously uninterrupted low-latency connections, DTN protocols were designed to accommodate long transmission delays, intermittent connectivity, and high error rates. Nonetheless, at the recent IETF meeting, a growing contingent has coalesced around using IP architecture centered on QUIC for deep space.

\textit{Lunar.} Lunar communications can be viewed as a middle ground for integrating near earth and deep-space solutions. Lunar communications typically have round-trip times of 5 to 14 seconds \cite{parisi_effects_2023}. In addition to distance, network designs must also account for lunar orbiting nodes being periodically unreachable due to being blocked by their orbit around the moon. Lunanet, a NASA and ESA project aimed at providing cis-lunar communications, approaches lunar communications issues with a hybrid architecture involving use of IP, 3PP, and DTN protocols \cite{noauthor_draft_2023}. 

%The chasm between space engineering and cryptography contributes to the split between using DTN protocols or IP protocols for space. The aforementioned nuances of the space environment are often overlooked by or not voiced to cryptographers designing security protocols. On the other hand, space engineers in both camps may overlook or not understand the buyer-beware fine print associated with security protocols designed by cryptographers (i.e. using a nail where a screw should be). Ultimately, both camps need to evaluate proper key establishment methods that can withstand the expansion of space communications. 

%Whether it's an IP stack, DTN stack, or a hybrid approach issues remain in bridging E2E secure communications for space applications. For security, DTN uses BPSec while an IP stack uses QUIC with TLS built-in. We show that neither of these are satisfactory due to security weaknesses and incompatibility with space environments and propose an alternative using MLS as a basis for key agreement. 

In both DTN and IP stack networking approaches, appropriate key establishment methods need to be selected. 
The following two subsections summarize the current security approaches for each. We then discuss the security considerations of these and current gaps.

\subsection{DTN Stacks with BPSec Security}

% What BPSec is/aims to achieve and its dependencies, how it works and what it was built for
The Bundle Protocol version 7 (BPv7 or simply BP)\cite{rfc9171} is a store-and-forward application layer protocol with integrated transport\footnote{Torgerson et. al. in \cite{rfc4838} refer to BP as existing in the ``bundle layer'', but we use classification by \cite{rfc9171} for clarity and familiarity.} currently deployed as a DTN component by NASA in the DTN stack \cite{manning_frequently_2023}. A bundle is comprised of a \textit{primary block} containing necessary header information, a \textit{payload block}, and \textit{extension blocks}. Bundles are passed hop-by-hop among source, intermediate, and destination nodes, operating on a best-effort store-and-forward basis. This may include \emph{convergence-layer protocols} (i.e. DTN adapted transport protocols) for end-to-end bundle delivery assurance. As such, BP is compatible with both DTN-specific transport protocols and  and TCP/IP based protocols 
(through their respective convergence-layer adapters such as  
QUICL). Security of the BP blocks is provided through the \emph{BPSec} protocol \cite{rfc9172}. 

% to store-and-forward bundles hop-by-hop between the source and destination, relying on \emph{convergence-layer protocols} (i.e. DTN adapted transport protocols) for end-to-end bundle delivery assurance. As such, BP is compatible with both DTN specific transport protocols
% % such as Licklider Transmission Protocol (LTP) or
% and TCP/IP based protocols 
% % TCP, UDP, and QUIC protocols 
% (through their respective convergence-layer adapters such as  
% % protocols TCP-CLA, UDP-CLA, and 
% QUICL). 

BPSec is a secure channel protocol that provides confidentiality and integrity. 
% protections to the blocks that comprise a bundle \cite{rfc9172} within BP. BPSec is 
Unlike a traditional secure channel, BPSec relies on shared keys across source, intermediate, and destination nodes. This is to enable flexibility; 
intermediate nodes are expected to process and potentially decrypt and act on bundles as part of the store-and-forward paradigm. Such functionality can enable intermediate node to act as network gateways or proxies, and even discard blocks. Supporting shared keys across all nodes raises questions about key establishment and maintenance, however. 
% This may include decrypting and discarding blocks as may be done by intermediate network gateways or proxies.
%\xt{DHTW25 (our paper)} characterize BPSec as a Flexible Secure Channel (FSC) and prove its security under a passive adversary model. 
BPSec, unlike traditional secure channels (i.e. IPSec), does not perform key establishment itself but instead relies on so-called out-of-band key establishment, i.e., pre-shared symmetric keys (PSKs)
under the default \emph{security context}, i.e., default mode 
% Default Security Context 
\cite{rfc9173}. PSKs are used for integrity protection in Message Authentication Codes (MAC) and confidentiality protection in  authenticated encryption with associated data (AEAD). 
Alternative ciphersuites and parameter sets may also be used for BPSec depending on 
the \emph{security context} 
supported by participating nodes. This allows for the potential introduction of
% which may include the use of 
asymmetric cryptographic methods and assignment as new security contexts \cite{ietf-dtn-bpsec-cose-05}.

\subsection{TCP/IP Stacks with QUIC Security}
%What QUIC is and how it work with the TLS handshake, what it aims to achieve and what it was built for
%\xt{TODO: Make the case that QUIC may be suitable for certain low-latency high availability space applications}

In the past, transport protocols such as TCP were defined separately from key establishment protocols, e.g., Transport Layer Security (TLS)~\cite{rfc8446}, leading to inefficiencies in connection setup and specifically, round-trips.
% between peers who wish to securely communicate. 
% Google's development of 
QUIC was developed as an ``all-in-one'' alternative approach, combining a UDP-based transport protocol with key establishment and secure channel protocols. By combining the two into a single protocol, various optimizations were possible. Key establishment in QUIC is based on the TLS Handshake protocol -- which it separates from the TLS Record Layer and Alert protocols (used in  TLS for the secure channel phase)~\cite{rfc9000}. It instead incorporates the TLS Handshake with a customized stack to reduce the inefficiencies in connection establishment. 
% This allowed Google Chrome, which was the first to incorporate QUIC, to dominate the internet browser space with markedly faster internet connections. 
% The built-in TLS allows two entities to reach a shared secret across an untrusted channel by using a Diffie-Hellman key exchange. 
Since the TLS Handshake provides entity authentication (either mutual or unilateral authentication) by leveraging public key infrastructure certificates, so too does QUIC. The symmetric keys derived from the handshake are used with AEAD for the QUIC secure channel. The QUIC secure channel does not use the TLS Record Layer protocol, however, instead opting for a separate customization in framing format. 
% While  TLS refers to the secure channel phase using AEAD as the Record Layer protocol, QUIC does not have similar dedicated terminology. 
Since reference terminology will be useful in what follows and QUIC does not define terminology for its components, we will abuse TLS terminology and refer to the components of the TLS-based QUIC Handshake as QUIC-HS and the subsequent QUIC secure channel as QUIC-RL (i.e., ``QUIC Record Layer'').  

QUIC's use of UDP instead of TCP has made it more appealing than other transport layer options for space applications, as latency is a concern. 
Under reliable network settings, QUIC performs better than other interactive IP protocols due to fewer round-trips 
% and thus are able to send authenticated encrypted data sooner than protocols like TLS over TCP 
\cite{shreedhar2021evaluating} since it removes the need to for a separate TCP SYN/ACK protocol initiation before the security protocol starts. 
While round trip reductions have improved the latency costs, QUIC is still a session-based protocol and the costs are not optimally minimized, as will be discussed in Section~\ref{sec:gaps}.
In space settings, any potential for latency reduction is critically important for consideration. 
% it especially appealing for certain space applications wishing to take advantage of an IP stack.  
% These entity authentication and key exchange functions provided by TLS are key components of the QUIC Handshake (QUIC-HS) which provides the symmetric key(s) used for authenticated encryption (AE) for QUIC streams by the QUIC Record Layer (QUIC-RL).  
%
% \vspace{-.1cm}

\subsection{Gaps and Issues}\label{sec:gaps}

% \subsubsection
\textit{BPSec.}
Generally, reliance on PSKs  in BPSec limits both security and functionality. First, the a static PSK incurs an increased vulnerability window, i.e., the compromise of the PSK reveals all communications in the past, present, and future that have been protected by that key. In cryptographic terms, it lacks \emph{forward secrecy} \cite{Boyd_Mathuria_Stebila_2020} (protection of past data in the event of compromise) and \emph{post-compromise security} \cite{7536374} (`self-healing' of the connection and protection of future data in the event of compromise, under certain conditions). 
Second, it has limited flexibility as it is difficult to add or remove access (the PSK must be distributed to all nodes through some pre-coordination). In cases of manual PSK installation this is particularly problematic. The lack of interoperability impacts federation potential, while no formal protocol definition for distribution of PSKs increases management complexity for devices and the system over time.
% it is impossible for outsiders to securely communicate with an existing security group without possessing the same PSK(s): a detriment to federation. While resiliency is a godsend for payload managers who get to fly a mission past its design lifespan for decades, they are a bane for PSK based security. Supporting legacy systems that use PSK(s) both limit key sizes and weakens the entire security posture as the network grows and keys are discarded to avoid reuse.

% \subsubsection{QUIC}
\textit{QUIC.} 
As with TLS, QUIC was designed for the interactive client-server model. It assumes synchronous, session-based connections between two parties where data is in the clear on both ends. 
 QUIC-HS is by default stateless, designed for internet-style use cases where tens of thousands of clients could easily be connecting to a single server per second \cite{yagna_pushy_2024}. Thus, QUIC executes full handshakes to establish each new connection. This requirement of interactive  connection establishment presents an issue in environments where low-latency is a concern. While this model is ideal for one-to-one or many-to-one internet applications such as web-browsing or content streaming,it is incongruent with the space setting. 

 Specifically, high latency and intermittent link availability directly makes session-based protocols sub-optimal; if a link has to be reestablished, an entire new handshake is required. Even with the round trip improvements in QUIC over TLS, this still costs a full round trip. With intermittent link availability, the cryptographic overhead of a session-based key establishment protocol used to create a secure channel
 becomes a self-imposed denial of service. 
Notably, a stateless protocol such as this is not even inherently required for most space settings; unlike with web traffic where it can be problematic to hold state for tens of thousands clients that may or may not reconnect to a server, the space nodes that connect to each other are relatively few in number.\footnote{Relatively few space devices and the potential for stateful connections is the very reason that  PSKs could even be considered in the BPSec context. 
} 
 
QUIC does support an option for stateful connections using \emph{session resumption} and \emph{0-RTT}, but sacrifices security for efficiency in such instances by reusing the previous connection state. The QUIC designers have advised disabling these features due to privacy concerns 
 % (i.e. session correlation) 
 and replay vulnerabilities 
 % via storage of TLS session tickets and address validation tokens 
 \cite{rfc9001}, making them ill-advised as options for solving the latency and handshake overhead challenges in space. Despite these warnings by the QUIC designers, early testing on use of QUIC in space systems have specifically looked at the 0-RTT option given normal latency concerns under secure QUIC session establishment~\cite{8450388}.
 \label{fn:statelessQuic}
 QUIC also does not offer post-compromise security once keys are lost, including in its session-resumption option. 
 Furthermore, attempts at 

% While this model is ideal for one-to-one or many-to-one internet applications such as web-browsing or content streaming,  
% under traditional network assumptions (low latency continuous links) 
% it is incongruent with the space setting. 
% High latency and intermittent links make any session 
% 
% it is not well suited for DTN settings (high latency intermittent links) \cite{6533337} nor many-to-many group settings. 
% Specifically, high latency and intermittent link availability directly makes session-based protocols sub-optimal; if a link has to be reestablished, an entire new handshake is required. Even with the improvements of QUIC, this still costs a full round trip. With intermittent link availability, the cryptographic overhead of a session-based key establishment protocol used to create a secure channel
% : the cryptographic overhead 
% becomes a self-imposed denial of service. 
% Notably, a stateless protocol is not even required for most space settings; unlike with web traffic where it can be problematic to hold state for millions of clients that may or may not reconnect to a server, the space nodes that connect to each other are relatively few in number.\footnote{Relatively few space devices and the potential for stateful connections is the very reason that  PSKs could even be considered in the BPSec context. 
% } 

Furthermore, the one-to-one QUIC connections  scale poorly ($\bigO(n^2)$) for groups of devices --  a space networking scenario that is anticipated by the space development community as evidenced by the design decisions of BPSec. 

\iffalse

\xt{remove? What-abouts and what-ifs on the key phase bit being inappropriate for large groups} 
Additionally, the alternating bit used in key updates by QUIC-HS for two-party settings is unsuitable for group settings to update a shared key used by many. In a two-part case, the key-phase bit is flipped to indicate use of a new key derived using a key derivation function (KDF) of the old key which is easily detected and synced. This mechanism was not designed for a multi-party case in which every member would need to agree on the bit flip which causes race conditions and/or additional overhead to manage.
\fi 
% why QUIC  is bad and not suitable for space, why BPSec is incomplete
%\input{2-RelatedWorks}
\section{The Space Security Environment}

%Why stateless is bad, handshakes are bad, delays and potential disruptions

The above issues highlight necessary directions for improving for key establishment designs relative to space settings. Such a protocol should 1) reduce round-trips to the maximum extent possible, to better support cases where latency is an issue.  Such a protocol is also, notably, 2) not required to be stateless, as unlike Internet use cases there is not a high risk of exhaustion from tens of thousands of different connection establishments. It is therefore possible to utilize stateful protocols that may offer better security or functionality benefits.\footnote{Note that a stateful protocols implies a typical ability to maintain state and not an inability to recover from state loss.} 
Such a protocol should 3) have the capacity to refresh keys at regular or predetermined intervals for forward secrecy and post-compromise security. It should 4) ideally be scalable to groups of devices (for BP), where there is the potential to insert further spacecraft into the communication, and 5) should support asynchronicity (for BP). 
% 
% Key agreement and entity authentication are cornerstones of modern secure applications as enablers of secure channels like QUIC-RL and BPsec. 
% As mentioned in Section~\ref{sec:gaps}, using a \emph{statically stateful} provisioned shared key establishment method like PSKs is too rigid and too unforgiving in the face of compromise. 
% On the other hand, using a \textit{synchronous} handshake-based \emph{stateless} key establishment protocol can stall secure channel setup in harsh space environments and at scale. 
What is most notable here is that the design space expands when stateful protocols are considered. 
Because connections in space are orders of magnitude longer-lived in terms of paired communicating identities (potentially decades) \cite{werner_how_2018} and orders of magnitude fewer than internet connections, ad-hoc state updates are advantageous 
% it is advantageous to use messages for state management rather than 
over handshakes: it takes fewer messages to maintain and/or update state than it does to complete handshakes to obtain fresh keys.
% for any given size group. 
% We now look to \emph{stateful asynchronous} protocols for a solution to key establishment for space environments: \textit{continuous key agreement} protocols. 

%Moreover, advanced properties such as forward secrecy (FS) and post compromise security (PCS) which guarantee data security before (FS) and after a compromise (PCS) have also become standard such as with Transport Layer Security (TLS) 1.3 and ratcheting protocols like Signal and Messaging Layer Security (MLS). 

% Why it is possible to handle stateful protocols in space better that internet connections - cite sources on number of separate connections / min or /sec in internet. 

\subsection{Solutions}
%CKA and its advantages as a general concept
Among stateful protocols with reduced latency, 
the category of continuous key agreement (CKA) protocols, a.k.a. ratcheted protocols, is salient. Here participants establish and maintain a shared cryptographic state which can be updated over time with fresh keying material. Instead of repetitious session establishment as in QUIC, CKAs establish a session once, and update keys as needed -- this enables no additional round-trip overhead or indeed even bandwidth costs when new transmission connections are started. The stateful protocol can instead send encrypted data immediately, without transmitting certificates or other keying material, with instead key updates taking place when appropriate as determined by the application. This is both an efficiency and security upgrade from stateless protocols like QUIC-HS/TLS. Furthermore, it does not inhibit support for cases of state loss, as a new session can be established, but rather lowers latency in all other cases. 
% can be generated dynamically and asynchronously. 

Additionally, many CKAs are asynchronous protocols, making them well-suited for the relay context such as BP requires, where
% ensure that participants can securely communicate to their partners even if their 
communication partners may be dynamically unavailable or transmissions pass through intermediate nodes.
% unlike their synchronous handshake counterparts.
% Being dynamically stateful, CKAs are able to generate fresh keys between parties without handshakes every time: both an efficiency and security upgrade from stateless protocols like QUIC-HS/TLS. 
% 
% 
% 
CKAs are also usually designed for forward secrecy and post-compromise security protections, with currently deployed examples including the Signal protocol~\cite{cohn2020formal},  
% This protects communications against key compromise that would otherwise expose all past and future encrypted communications. 
% The Signal Double Ratchet protocol, 
WhatsApp Sender Keys protocol~\cite{AC:BalColGaj23}, and the Messaging Layer Security (MLS) protocol~\cite{C:AlwJosMul22}.
% are examples of currently deployed stateful asynchronous CKA protocols on the internet. 
% 
While CKA deployments have historically been in messaging applications, there have been increasingly  other  applications  in disrupted, relay, or high-latency settings~\cite{webex,drones8050200}-- making them unsurprisingly suited for space. 

%MLS as a standardized CKA concete option
Among CKAs, MLS features as an IETF standardized, open protocol 
% which provides advantages not afforded to other protocols like the Signal Double Ratchet or WhatsApp Sender Key protocols
\cite{rfc9420}. IETF standardization 
has been a feature for other space system deployments (both BPSec and QUIC being standardized in this way), and could further enable 
% protocols have undergone thorough review and considered ready for internet use (and are often already deployed) making them easier to adopt 
 adoption by organizations like the CCSDS and its members (e.g. NASA and ESA). MLS can also be deployed with post-quantum security~\cite{mahy-mls-xwing-00,hale-mls-combiner-01}, including efficient and flexible post-quantum options for amortizing the computational cost of post-quantum algorithms~\cite{hale-mls-combiner-01}.
% Furthermore, as a standardized internet protocol, MLS is more studied and scrutinized by public and private sector stakeholders from across the world for security and interoperability. MLS is one of the few concrete options that can be deployed immediately for space key agreement. 

\subsection{Functionality and Security Comparisons}

At a high level, MLS is not only a CKA, but a multi-party CKA, which supports goal (4) for scalability. It was designed to support scalable connections for multiple parties without the need for pairwise handshakes, or indeed all parties being connected simultaneously.  
% allows multiple parties to generate a shared key without performing pairwise-handshakes. 
% MLS continuously evolves the key (i.e. in an indistinguishable-from-random manner) without all parties being online, and 
It scales logarithmically with the size of a group ($\bigO(n\log(n))$ vs $\bigO(n^2)$ for pairwise QUIC handshakes), and supports operations to \emph{update} keying material, \emph{add} new group members, and \emph{remove} existing members.   
% It does all this
% by storing the group state (used to compute a key) as a tree that can be updated by group members (represented as leaf nodes). Group operations are sent as MLS messages to \textit{update} the group state, \textit{add} new members, and \textit{remove} existing members. 
% 
% 
%To facilitate group operations, MLS requires supporting components: an Authentication Service (AS) to associate identities to public keys and a delivery service (DS) to that ensures the receipt of MLS messages by participating nodes. 
% MLS is flexible in its architectural requirements to 
MLS supports client-server models and pairwise connections where desired (e.g., analogous to QUIC), as well as  decentralized models (similar to BPSec). 
% to which can be relevant for many space applications. 

% Comparison of MLS to current approaches
\iffalse
\xt{consider removing due to redundancy in points made} Ultimately, MLS fulfills the gaps left by PSKs and QUIC-HS/TLS have in key establishment for space networks. Mainly, static PSKs have no recourse against key compromise whereas MLS achieves FS and PCS with key updates. QUIC-HS/TLS must set up new sessions using interactive handshakes with online participants and to attain similar protections.  
\fi 
%security guarantees

% functionality 

\section{CKA Integration with QUIC and BPSec}

\begin{figure}
    \centering
    \includegraphics[width=\linewidth]{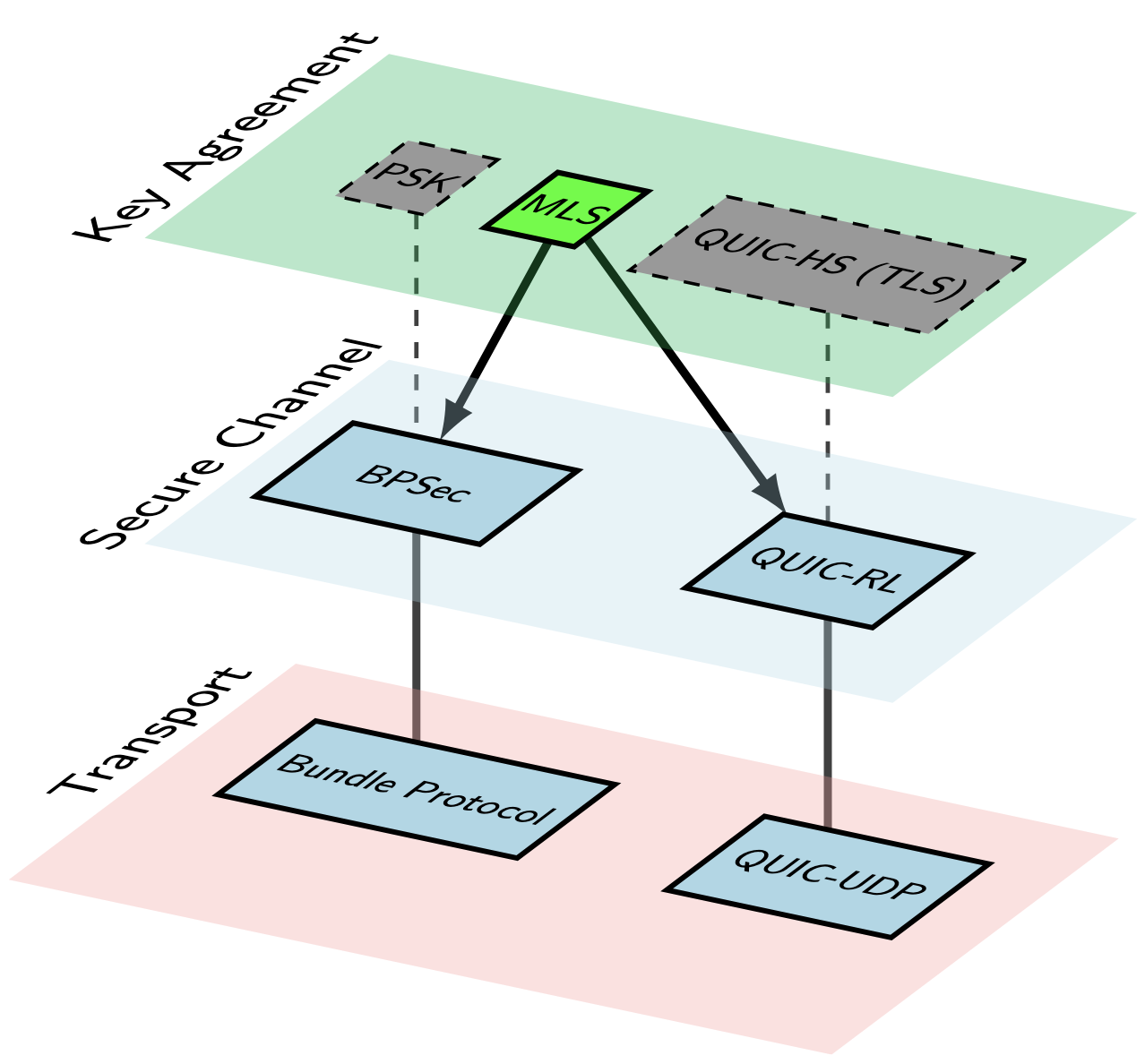}
    % \includestandalone[]{figs/MLSLayerPaths}
    \caption{MLS as a Key Establishment for BPSec or QUIC.  As shown, MLS provides the key (replacing the PSK)  in BPSec. It can also replace the TLS handshake in QUIC, with resultant keys usable directly in the QUIC ``Record Layer''. 
    % The sub-functions of QUIC have been abstracted into separate transport, secure channel and key agreement layers for comparison purposes.
    Solid edges represent the envisioned stack while dashed edges represent current pathways. 
    % 
    % existing paths and dashed edges represent proposed integration paths. 
    Convergence Layer adapters can be used to integrate QUIC  with BP, if desired, and are not shown.
    % \xt{is it confusing to have the PSK edges since PSK usage in both MLS and TLS differ quite a lot and come with serious vulnerability caveats?}
    }
    \label{fig:MLSHighLevel}
\end{figure}

%Integrated design of transport and security protocols provide bespoke solutions to a goal but when the problem space is changed from terrestrial to space networks or stateless to statefull security then the design must be carved out into pieces to reassess. 
Since a CKA is a key establishment protocol it offers the possibility for smooth integration in the 
 existing space network architectural landscape.
 % provides flexibility for the incorporation of MLS in space communications. 
 Here we describe at a high level how MLS, as an example CKA, can be incorporated into the both the DTN stack and the IP stack approaches
 %, as illustrated in 
 (Figure \ref{fig:MLSHighLevel}). 

\subsection{MLS as Key Establishment for BPSec}
As noted previously, BPSec assumes out-of-band key agreement, typically assuming a PSK. 
MLS outputs keys that is indistinguishable from random (a standard cryptographic assumption for key establishment)~\cite{C:AlwJosMul22}. Such keys can be used directly for the BPSec  channel wherever a shared key is required. 

% In practical implementation terms

 % As a key agreement protocol, existing MLS implementations\footnote{\url{https://github.com/mlswg/mls-implementations}} can be reused for BPSec key agreement so long as the MLS architectural components are implemented. 

In further practical implementation considerations, MLS implementations~\cite{mlsimp} use \textit{pre-key bundles} which contain public keys. Distribution of these public keys can be fulfilled through DTN-tailored Public Key Infrastructure protocols such as the Delay Tolerant Key Administration (DTKA) protocol or KeySpace~\cite{burleigh-dtnwg-dtka-02, smailes_keyspace_2024}.\footnote{PSKs \emph{can} be used to bootstrap MLS key agreement instead of public keys, but it is non-standard practice and comes with security considerations\cite{rfc9420}. Normal MLS initial key establishment is advisable.}  
% The distribution service function may be fulfilled in a few other ways. 
In a pure DTN stack, the Licklider~\cite{rfc5326} protocol, which provides reliable data transport using scheduled transmissions and quality of service mechanics 
% of red (reliable) and green (unreliable) data may be used as a DS 
could provide the underlying transport delivery service assurance.
% 
% 
% In a hybrid IP stack using a QUIC Convergence Layer Adapter \cite{QUICL}, QUIC transport meets functional requirements of a MLS delivery service, with MLS performing the QUIC-HS role. 
% 
% 
%As a third alternative, peer-to-peer or distributed protocols may also be used to perform DS roles through seeding and downloading the MLS group state. 
%These approaches exemplify strongly consistent DS mechanisms tailored for DTNs, as they ensure the timely and ordered delivery of MLS messages. 
%whereas the latter approach represents an eventually consistent DS model, relying on peer-driven decentralized consensus to establish the group state.

% The first two approaches represent DTN tailored strongly consistent DS approaches due to their guarantees on timely and ordered delivery of MLS messages while the latter represents an eventually consistent DS due to its reliance on peers to reach a decentralized consensus on group state.  

\subsection{MLS as Key Establishment for QUIC}

We present the novel recommendation for incorporating the MLS key establishment and large QUIC protocol. Doing so within QUIC is 
% Within QUIC, incorporating MLS is 
notionally even more straightforward than in BPSec.
Namely, the current QUIC-HS is a TLS-based handshake, which itself has been cryptographically analyzed for producing indistinguishable-from-random keys~\cite{10.1145/2976749.2978325}, which are then used in the secure channel phase (QUIC-RL). 
% Thus, in security terms, replacing QUIC-HS with MLS results in 
% 
% since can be easily adapted in the QUIC-HS role. 
% Recall that QUIC-RL requires the following functions from QUIC-HS: an authenticated exchange of keys, transport parameters, and application protocols \cite{rfc9000}. 
Since MLS also produces indistinguishable-from-random keys~\cite{10.1007/978-3-030-56784-2_9}, not only is the functional replacement imply that these keys can be used directly in the QUIC secure channel phase (QUIC-RL), but that security analyses are composable. 

Furthermore, MLS provides an authenticated key exchange which can be continuously evolved, either ad-hoc or on a pre-determined schedule. This removes the security need for a round trip when previously connected devices reconnect, and removes  the session resumption risk of QUIC (see \cref{fn:statelessQuic}).  
% 
% satisfies these by setting up a new MLS group through a series of authenticated group addition messages. Then, with a shared group state, keys can be derived asynchronously which allows for transport parameters to be confidentially negotiated in an authenticated manner. 
MLS also supports authenticated negotiation of application protocols using the content advertisement extension~\cite{ietf-mls-extensions-04}. In turn, in contexts where QUIC is already considered for space applications, MLS does not introduce any changes
% architectural requirements are complimented by QUIC through reuse of 
certificate management but could, optionally, use lighter weight certificate management approaches due to the flexibility in ``identity'' definition~\cite{rfc9420}. 
Since MLS is a change and replacement for the key establishment phase only, any desired connection reliability benefits from 
% as an authentication service and utilizing 
QUIC's transport are preserved.

\section{Conclusion}

The disconnect between 
% The development chasm between 
the space engineering and cryptography fields risks contributing to a design of space networking that is not suited for its purpose, as nuances in either field are not always well communicated to the other. 
Functionality requirements in space are different than in Internet connections and change the types of security protocols that should be considered. Failure to choose security protocols that meet the the use case can risk long-term sub-optimal functionality or security vulnerabilities. 
% On the other side, cryptographic developments that reach beyond ``traditional'' Internet options and may be suitable for a variety of purposes are not always presented 
% 
% On one hand, the nuances of space environment are often overlooked by or not voiced to cryptographers designing security protocols. On the other, space engineers may overlook or not understand the buyer-beware fine print associated with security protocols designed by cryptographers (i.e. using a nail where a screw should be). 
This work outlines existing conflicts among methods for key establishment in space communications and 
we propose a path towards both functional and security improvements that meets various networking stack approaches. 
% proposes to use MLS to address security and efficiency concerns. 
Ultimately, progression in space has always required close collaboration across many disciplines, and 
% development of secure networking is no different. 
% need to evaluate proper 
that continues to apply with the expansion of security in space. 
% to the emergence and application of 
% key establishment methods that can withstand the expansion of space communications.

% QUIC as a HS + RL
% How TLS was HS + RL and QUIC took took HS out and added its own RL
% How MLS can replace the HS while keeping QUIC RL

% % \section{Bibliography}
% Citing papers is done in the usual way using BibTeX or \texttt{biblatex}
% commands. For example: the RSA paper~\cite{RSA78}.

% It is highly encouraged to use CryptoBib from \url{https://cryptobib.di.ens.fr}
% \newpage
\bibliographystyle{ieeetr}
% This sample uses bibtex rather than biblatex.
\bibliography{refs, abbrev3, crypto,biblio}

% NOTES
% - Download abbrev3.bib and crypto.bib from https://cryptobib.di.ens.fr/
% - Use bilbio.bib for additional references not in the cryptobib database.
%   If possible, take them from DBLP.

\end{document}